\documentclass{jfm}
\pdfoutput=1

\usepackage{graphicx}
\usepackage{newtxtext}
\usepackage{newtxmath}
\usepackage{natbib}
\usepackage{hyperref}
\usepackage{commath}
\usepackage{bm}
\usepackage{xcolor}
\usepackage[justification=justified,width=\textwidth]{caption}
\hypersetup{
    colorlinks = true,
    urlcolor   = blue,
    citecolor  = black,
}

\newcommand{\RomanNumeralCaps}[1]
\linenumbers

\title{Temporal large-scale intermittency and its impact on the statistics of turbulence}

\author{Lukas Bentkamp\aff{1}\corresp{\email{lukas.bentkamp@uni-bayreuth.de}}
 \and Michael Wilczek\aff{1}
 \corresp{\email{michael.wilczek@uni-bayreuth.de}}}

\affiliation{\aff{1}Theoretical Physics I, University of Bayreuth, Universit\"atsstra\ss e\ 30, 95447 Bayreuth, Germany}

\begin{document}
\maketitle

\begin{abstract}
  Turbulent flows in three dimensions are characterized by the transport of energy from large to small scales through the energy cascade. Since the small scales are the result of the nonlinear dynamics across the scales, they are often thought of as universal and independent of the large scales. However, as famously remarked by Landau, sufficiently slow variations of the large scales should nonetheless be expected to impact small-scale statistics. Such variations, often termed large-scale intermittency, are pervasive in experiments and even in simulations, while differing from flow to flow. 
  Here, we evaluate the impact of temporal large-scale fluctuations on velocity, vorticity, and acceleration statistics by introducing controlled sinusoidal variations of the energy injection rate into direct numerical simulations of turbulence. We find that slow variations can have a strong impact on flow statistics, raising the flatness of the considered quantities. We discern three contributions to the increased flatness, which we model by superpositions of statistically stationary flows. Overall, our work demonstrates how large-scale intermittency needs to be taken into account in order to ensure comparability of statistical results in turbulence.
\end{abstract}

\section{Introduction}

In developing statistical theories of turbulence, a certain degree of universality is  often assumed.
However, even the precise values of supposed universal constants such as the Kolmogorov constant for the energy spectrum are still up for debate~\citep{SreenivasanPF1995,YeungPRE1997,DonzisJFM2010,IshiharaPRF2016}. 
In part, this may be because very different types of flows (including those from simulations) are attempted to be described under one umbrella.
While it is agreed that the large scales of turbulence differ between flow types, there is still the expectation
that the small scales of turbulence approach some universal statistical state at large Reynolds numbers, as first hypothesized by Kolmogorov in his classical theory of turbulence~(hereafter K41, see \citet{kolmogorov1941a}). In his theory, the central role is played by the mean dissipation rate $\langle \varepsilon \rangle$ that determines the average energy flux from large scales to small scales. If we assume that the value of this average energy flux is the only information that propagates through the cascade dynamics from large to small scales, then the small-scale statistics have to take the universal form dictated by the K41 theory.
However, the idealized assumptions of the K41 theory are not quite correct. In particular, it turned out that there is a build-up of fluctuations around the average flux of energy toward smaller scales~\citep{CeruttiPoF1998,BuzzicottiJT2018,YasudaJFM2018}, leading to extreme fluctuations at the smallest scales. This is known as small-scale intermittency~\citep{SreenivasanARFM1997}. While this observation contradicts the K41 theory, it does still allow for universality of small-scale statistics for a given Reynolds number.

Complementarily, a limitation of applicability of the K41 theory was brought up shortly after publication by Landau, who remarked that sufficiently slow (or sufficiently spatially extended) flow variations cannot be expected to be filtered out by the cascade dynamics and will lead to alterations of small-scale statistics~\citep{Landau2013,KraichnanJFM1974,Frisch1995}. This is now known as large-scale intermittency (see \citet{chien2013} for a discussion of the terminology). A way to rationalize his criticism is to assert that turbulent flows have finite spatio-temporal correlations. Therefore, the small-scale statistics at a certain point in space and time can depend on the flow only within a certain spatio-temporal region. Any variations on larger scales will generate dissimilar, statistically independent domains of the flow. Aggregated statistics over these domains will necessarily depend on the large-scale variations. Contrary to small-scale intermittency, the large-scale variations are expected to differ from flow to flow.

Technically, such large-scale variations could be considered in conflict with Kolmogorov's assumption of local isotropy~\citep{kolmogorov1941a}, implying that K41 theory does not apply here. While one could therefore exclude such very large-scale variations from the discussion of universality~\citep{KraichnanJFM1974}, it is important to note that they are not uncommon in real-world turbulent flows: For example, large-scale variations in atmospheric turbulence were observed to enhance the effect of small-scale intermittency~\citep{BoettcherBM2003,MuschinskiJFM2004,BoettcherSERRA2007,FeracoE2021}. Similarly, many experimental set-ups for turbulence produce large-scale fluctuations~\citep{MouriPF2006,mouri2009,blum2011}, such as von K\'arm\'an flow~\citep{voth2002,MordantNJP2004,LawsonEF2014}. Certain types of simulated flows can also have emerging large-scale variations such as Kolmogorov flow~\citep{BorueJFM1996,GotoFDR2016,LalescuPRR2021}, or they can feature shifting turbulent/non-turbulent interfaces such as turbulent jet flows~\citep{GaudingPotCI2021,GaudingJFM2021}. 
By contrast, many numerical turbulence simulations are designed to display a minimum of large-scale fluctuations, e.g., by enforcing a fixed total energy~\citep{ishihara2007}. 
In practice, spatial or temporal averages are often performed to improve sample sizes, but this may hide the presence (or absence) of the large-scale fluctuations.

\citet{kolmogorov1962} and \citet{oboukhov1962} partly addressed Landau's remark in their refined similarity hypotheses (see also~\citet{MouriPF2006} and \citet{chien2013}). Instead of basing their theory on the mean energy dissipation rate $\langle\varepsilon\rangle$, they postulated universal statistics on a length scale $r$ as a function of the energy dissipation spatially averaged over scale $r$. This theory constituted an improvement over the previous one by incorporating small-scale intermittency, acknowledging the fact that energy dissipation fluctuates strongly in space and time~\citep{MeneveauJFM1991}. However, the theory assumed universal log-normal large-scale fluctuations and did not consider external variations in the sense of Landau, which may depend on the flow type. 

In the literature, the impact of large-scale intermittency on statistics is not often discussed explicitly.
There are some parts of the literature speaking about external intermittency as the random switching between turbulent and non-turbulent flow, which can affect inertial-range properties and specifically higher-order statistics~\citep{KuznetsovJFM1992,MiPRE2001,GaudingPotCI2021,GaudingJFM2021}.
Here, we consider fully developed turbulence, and we understand large-scale intermittency as fluctuations on scales comparable to or larger than the integral scales. 
In order to assess the impact of such large-scale fluctuations, Monin and Yaglom proposed a two-state model explaining how the mixing of flow regions with different dissipation rates changes the coefficients of structure functions while retaining their scaling~\citep{Monin2013,DavidsonPT2005,chien2013}.

One approach to quantify the impact of large-scale fluctuations is to compute conditional statistics. It was found that in many cases, second-order structure functions conditional on the large-scale velocity indeed depend on these large scales~\citep{PraskovskyJFM1993,SreenivasanPRL1996,SreenivasanPoTPS1998,blum2010,blum2011,chien2013}. Some studies also considered conditional higher-order statistics in this context~\citep{PraskovskyJFM1993,blum2010,GaudingJFM2021,GaudingPotCI2021}.
\citet{BoettcherBM2003,BoettcherSERRA2007} found that the increment statistics of atmospheric turbulence matches better with laboratory data when conditioning on the mean wind speed and proposed an ensemble model based on this idea.
\citet{ShnappJFM2021} found that also the Lagrangian statistics in canopy flows are affected simultaneously by small-scale and large-scale intermittency.

Complementarily, introducing artificial variations of the large scales allows one to study their impact more systematically. \citet{KnebelEF2011} used an active grid to generate statistics resembling intermittent atmospheric data. \citet{chien2013} studied how the small-scale statistics of a flow between oscillating grids changes with different degrees of large-scale variations. We here adopt a similar approach by introducing artificial variations into direct numerical simulations. 
Given that the non-trivial behavior of higher-order statistics is a hallmark feature of turbulence, we focus on the effect of large-scale intermittency specifically on higher-order statistics at the example of the flatness and propose modeling strategies.

In a parallel line of research, temporal variations of the energy input (akin to the ones that we analyze in the present work) have been used to study the response and resonance behavior of turbulence. Given periodic driving, sometimes a resonant response of the flow can be observed~\citep{CadotJFM2003,vonderheydt2003a,vonderheydt2003b,kuczaj2006,kuczaj2008,cekli2010}. Some studies considered periodically kicked turbulence~\citep{lohse2000,jin2008}. It was also found that periodic driving can affect transport properties~(\citeauthor{jin2008}\ \citeyear{jin2008}; \citeauthor{BosCF2017}\ \citeyear{BosCF2017}; \citeauthor{YangFDR2019}\ \citeyear{YangFDR2019}; \citeauthor{yang2020}\ \citeyear{yang2020}). When driving turbulence with periodically varying shear, there is a critical frequency above which the turbulent flow can no longer be sustained~\citep{YuJFM2006,hamlington2009}. Furthermore, studying variations of the energy input can provide insight into the workings of the energy cascade~\citep{BosPF2007,FangJT2023}. While we here study a similar set-up with periodic driving, we differ from this line of research by taking the modulated driving as a model for large-scale intermittency and investigating its impact on time-aggregated, higher-order statistics compared to simulations without the modulation.

In this study, we ask how much and by which mechanisms the higher-order flow statistics are affected by slow fluctuations at the large scales. Then we ask whether these effects can be explained by models based on ensembles of statistically stationary flows. To address these questions, we introduce sinusoidal variations of the energy injection rate into direct numerical simulations of turbulence. While these variations are to be understood as a model for the large-scale fluctuations that can occur in various forms in any flow, we focus on sinusoidal injection-rate signals, which allow for a systematic investigation using periodic averaging. For different frequencies of the oscillations, we analyze the statistics of velocity, vorticity, and Lagrangian acceleration.
We observe an amplification of flatness, which we attribute to three different effects. The first and largest contribution comes from mixing distributions with varying width and can simply be estimated from the time series of mean energy and dissipation rate. The second contribution results from variations of higher-order flow statistics and can be described by an ensemble of stationary flows. A third contribution occurs at certain frequencies of the input, where we observe that time-averaged statistics are not simply a mix of statistics corresponding to the various injection rates. Instead, the second moment of the dissipation rate displays stronger excursions than expected. By considering the time series of both mean and variance of the dissipation rate, we construct an ensemble of stationary flows that can capture all three effects and thus accurately predict flatness values. Finally, we also explore how our results are affected by the Reynolds number.

\section{Direct numerical simulations} \label{sec:dns}
\begin{figure}
 \centerline{
   \includegraphics{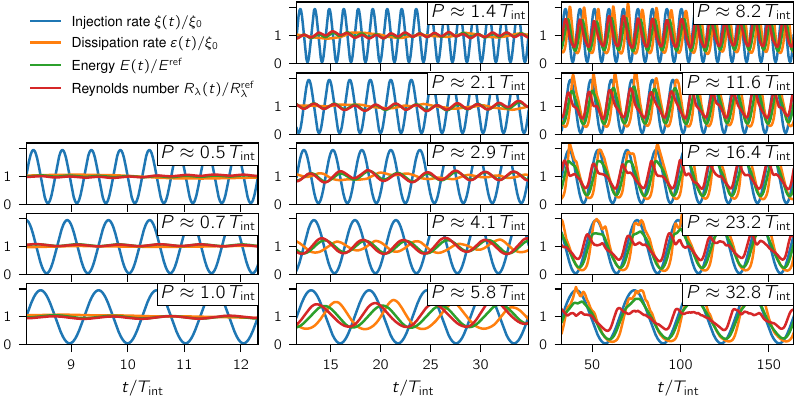}
 }
 \caption[width=\linewidth]{Time series of different flow characteristics in simulations with oscillating energy injection rate at period times $P$. All quantities are normalized by their average value from the reference simulation with injection rate $\xi(t) = \xi_0$, average energy $E^\mathrm{ref}$, and Reynolds number $R_{\lambda}^{\mathrm{ref}} \approx 103$. The time-dependent Taylor-scale Reynolds number is computed as $R_\lambda(t) = E(t) (20/(3\nu\varepsilon(t)))^{1/2}$ and thus peaks when energy is large compared to dissipation rate. The fact that the amplitude of the flow response decreases with the oscillation frequency shows that the flow acts as a low-pass filter. The code and post-processed data used to generate this figure can be explored at \url{https://www.cambridge.org/S0022112024007006/JFM-Notebooks/files/Figure-1.ipynb}.\label{fig:time_series}}
\end{figure}
We analyze data from direct numerical simulations (DNS) of the forced, incompressible Navier-Stokes equation in the vorticity formulation,
\begin{align}
  \partial_t \bm \omega + \bm u \cdot \nabla \bm \omega &= \bm\omega\cdot\nabla\bm u + \nu \nabla^2 \bm \omega + \bm f,
\end{align}
solved by a pseudo-spectral code~\citep{LalescuCPC2022}. Here, $\bm \omega(\bm x, t)$ is the vorticity field on a periodic domain~$\bm x \in [0, 2\pi]^3$ (discretized on $512^3$ grid points), $\bm u(\bm x, t)$ is the velocity field with $\nabla\cdot \bm u = 0$, $\nu$ is the kinematic viscosity, and $\bm f(\bm x, t)$ is the forcing.
The flow is forced on the large scales by amplifying a discrete band of Fourier modes $k \in [1.4, 2.3]$ (DNS units), imposing a prescribed energy injection rate $\xi(t)$ (specified below). Due to the forcing, the simulations are approximately statistically homogeneous and isotropic. The spatial resolution varies between $2.4 \leq k_M\eta \leq 6.4$ across the simulations, where $k_M$ is the maximum wavenumber, and $\eta$ is the Kolmogorov length scale. Each simulation lasts between $131\,T_\mathrm{int}$ and $262\,T_\mathrm{int}$ (where $T_\mathrm{int}$ is the integral time scale). In order for the statistics to be independent of the initial condition (which is a statistically stationary turbulent field), we use only data after a transient period of at least $8\, T_\mathrm{int}$.

The main focus of this study is on unsteady simulations with periodically varying injection rate,
\begin{align} \label{eq:injection_rate}
  \xi(t) &= \xi_0 + A_\xi \sin(2\pi t/ P),
\end{align}
where $\xi_0$ is the base injection rate, $A_\xi = 0.95\xi_0$ is the amplitude of the oscillations, and the period time $P$ is varied from $0.5\, T_\mathrm{int}$ up to $32.8\, T_\mathrm{int}$. Throughout this paper, all reference values such as the integral scales and the Kolmogorov scales are taken from a simulation with constant injection rate $\xi(t) = \xi_0$ at Taylor-scale Reynolds number $R_{\lambda}^\mathrm{ref} \approx 103$ (with the exception of the resolution criterion $k_M\eta$, for which $\eta$ is computed separately in each simulation).

In the course of this work, we will associate the statistics of the oscillating simulations with the statistics of simulations at various constant injection rates $\xi(t) = \alpha\xi_0$. After the transient, these simulations become approximately statistically stationary. An ensemble of 23 simulations on a $512^3$ grid with $\alpha$ linearly spaced between $0.05$ and $2.14$ ($R_\lambda$ between 60 and 120) allows us to cover all of the injection rates reached by the oscillating simulations. An extension of 3 simulations on a $1024^3$ grid with $\alpha \in \{4, 9, 18\}$ ($R_\lambda$ between 134 and 173, each lasting $41\, T_\mathrm{int}$) will help us in sections~\ref{sec:dynamical_effects} and \ref{sec:fluct_diss_ens} to model features of high-Reynolds-number turbulence that we observe in the oscillating simulations. A separate set of higher-Reynolds-number simulations is described and analyzed in section~\ref{sec:higher_Re} in order to evaluate the Reynolds number dependence of our results.

\section{Results}
\begin{figure}
  \centerline{
    \includegraphics{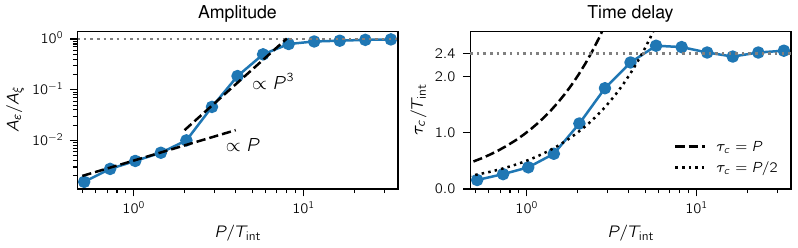}
  }
  \caption[width=\linewidth]{Amplitude $A_\varepsilon$ and time delay $\tau_c$ of the variations of dissipation rate, determined by finding the maximum point of the cross-correlation between injection rate and dissipation rate, $A_\xi A_\varepsilon/2 = \max_\tau \overline{\langle \xi(t-\tau) \varepsilon(t) \rangle} - \xi_0^2$ and $\tau_c = \mathrm{argmax}_\tau \overline{\langle \xi(t-\tau) \varepsilon(t) \rangle}$ with $0 \leq \tau \leq P$. Note that these formulas give the exact amplitude and phase shift for sinusoidal signals. The code and post-processed data used to generate this figure can be explored at \url{https://www.cambridge.org/S0022112024007006/JFM-Notebooks/files/Figure-2.ipynb}.
  \label{fig:cascade_time}}
\end{figure}

To illustrate the flow response in the oscillating simulations, figure~\ref{fig:time_series} shows time series of various flow characteristics. The energy and the dissipation rate resemble
the injection rate signal with a delay. The amplitude of their oscillations reduces at smaller period times, indicating that the energy cascade acts as a low-pass filter~\citep{CadotJFM2003,kuczaj2006,BosPF2007}. This filtering effect is analyzed in more detail in figure~\ref{fig:cascade_time}, where we show the relative amplitude and the time delay of the variations of the dissipation rate compared to the injection rate. Only for the slowest variations of the injection rate (largest~$P$) does the dissipation rate oscillate at the same amplitude. As soon as the variation time scales become comparable to the integral time of the flow, the amplitude drops. The decrease is consistent with the $P^1$ and $P^3$ scalings reported in the literature~\citep{kuczaj2006,BosPF2007}, where the transition threshold between these two regimes depends on the Reynolds number. Similarly, the time delay saturates at a well-defined cascade time of approximately $2.4\, T_\mathrm{int}$ only for the slowest oscillations. Due to the method that we use to measure the time delay $\tau_c$, it cannot be larger than the period time $P$. Instead, for small period times, we observe that it follows $P/2$ approximately, which corresponds to a phase shift of $\pi$ \citep[compare][]{kuczaj2006,BosPF2007}.
\begin{figure}
  \centerline{
    \includegraphics{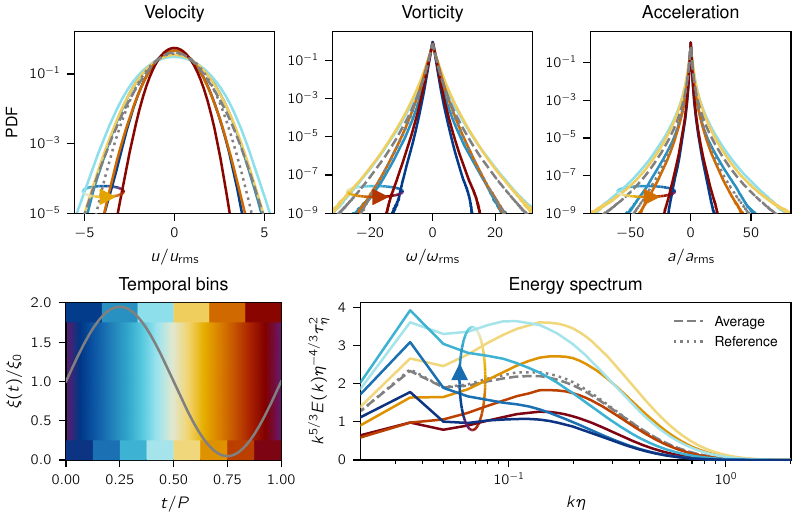}
  }
  \caption[width=\linewidth]{Periodically averaged statistics of the oscillating flow with period time $P \approx 8.2\,T_\mathrm{int}$. Top: The PDFs of velocity, vorticity, and acceleration. Here, $u_\mathrm{rms}$, $\omega_\mathrm{rms}$, and $a_\mathrm{rms}$ denote the root mean square of the velocity, vorticity, and acceleration, respectively, all computed from the reference simulation with $\xi(t) = \xi_0$. A directed circle indicates the arrow of time. Bottom left: Visualization of the color code for the temporal bins. For the computation of the PDFs, the oscillation period is split into $n=6$ equally spaced bins (top). For the spectra, $n=8$ temporal bins are used (bottom). The injection rate (grey) is shown for reference. Bottom right: Compensated instantaneous energy spectra. In the course of each period, the energy is first injected at small wavenumbers, then transported to small scales, and then dissipated. The code and post-processed data used to generate this figure can be explored at \url{https://www.cambridge.org/S0022112024007006/JFM-Notebooks/files/Figure-3.ipynb}. Animated versions of this figure can be found in the supplementary material (available at \url{https://doi.org/10.1017/jfm.2024.700}) as movie 1, movie 2, and movie 3 for the simulations with $P \approx 2.1\,T_\mathrm{int}$, $8.2\,T_\mathrm{int}$, and $32.8\,T_\mathrm{int}$, respectively. Movie 4 shows a three-dimensional visualization of the energy density in the simulation with $P \approx 8.2\,T_\mathrm{int}$.} \label{fig:breathing_pdfs}
\end{figure}

\subsection{Periodically averaged statistics}
We investigate three exemplary single-point flow statistics: components of velocity, vorticity, and Lagrangian acceleration, denoted by $u$, $\omega$, and $a$, respectively. Since we are specifically interested in the effect of the periodically oscillating energy injection rate, we introduce a periodic average $\langle \cdot \rangle$ that averages over all degrees of freedom apart from the phase of the oscillation. All other degrees of freedom, namely the spatial dependence, the three vector components, and full periods of the oscillations, are expected to be statistically equivalent and thus averaged over. Hence a periodically averaged quantity $\langle \cdot \rangle$ depends only on the phase $0\leq t\;\mathrm{mod}\;P < P$. We further increase the number of samples by aggregating samples of similar phase in linearly spaced temporal bins $0 = t_0 < \dots < t_n = P$ such that the $i$th bin is given by all samples with
\begin{align}
  t_{i-1} \leq t \;\mathrm{mod}\; P< t_i.
\end{align}
We choose a suitable number of bins $n$ (as indicated throughout the paper) depending on the need. For short period times, we reduce $n$ such that temporal bins are not shorter than $\tau_\eta/2$ (where $\tau_\eta$ is the Kolmogorov time scale). The corresponding phase-dependent probability density functions (PDFs) based on the periodic average are shown for $P \approx 8.2\, T_\mathrm{int}$ in figure~\ref{fig:breathing_pdfs} (top). While velocity distributions are generally close to Gaussian, vorticity and acceleration distributions display heavier tails. Over the period of the oscillations, the width of the PDFs varies strongly.
With the artificial oscillations, we imitate the effect of the large-scale fluctuations that may occur in various types of turbulent flows. If such temporal variations are simply averaged over, then this may change flow statistics compared to flows without such variations. For this study, we compare time-averaged statistics of oscillating flows, such as the distributions obtained as the mean of the oscillating PDFs in figure~\ref{fig:breathing_pdfs} (top, grey dashed lines), with statistics from the reference flow at stationary injection rate $\xi(t) = \xi_0$ (grey dotted lines). After characterizing the impact of the temporal variations, we pose the question of whether it can be modeled statistically as a mixture of an ensemble of stationary flows.

Another quantity that serves to illustrate the different states of the oscillating flows is the compensated energy spectrum shown in figure~\ref{fig:breathing_pdfs} (bottom right). Over the first half of an oscillation period, energy is injected at small wavenumbers~$k$. During this time, the compensated energy spectrum peaks at those small wavenumbers. In the second half of the oscillation period, the energy cascades toward larger wavenumbers while the energy injection is low. During this time, the compensated energy spectrum peaks at large wavenumbers and decays. 

\subsection{Impact of oscillations on time-averaged statistics}
In order to quantify the probability of extreme events, we consider the total (i.e. time-aggregated) ($k$th-order hyper-)flatness, defined as
\begin{align} \label{eq:total_flatness}
  F_{X,k}^\text{tot} &= \frac{\overline{\langle X(t)^k \rangle}}{\overline{\langle X(t)^2 \rangle}^{k/2}}
\end{align}
for the three quantities $X = u, \omega, a$ (components of velocity, vorticity, and acceleration) and even $k\geq 4$. Note that all three quantities have zero mean. The overbar denotes temporal averaging:
\begin{align}
  \overline{f(t)} &= \frac{1}{P} \int_0^P \dif t\, f(t).
\end{align}
Since the periodic average $\langle \cdot \rangle$ depends only on the phase of the oscillation $t\;\mathrm{mod}\;P$, the temporal average ranges only from $0$ to $P$.
For simplicity, we do not write the dependence on space and on vector component indices, which are averaged over by the periodic ensemble average $\langle \cdot \rangle$ (described above). We will consider the cases of regular flatness (or kurtosis) with $k=4$ and the hyper-flatness with $k=6$. Since the high-order moment puts emphasis on the tails of distributions, the flatness is large whenever the tails have comparably high probability. Since the flatness is non-dimensional, it does not change when linearly rescaling the PDF, e.g., Gaussian distributions always have a regular flatness 3 and a 6th-order hyper-flatness 15.

\begin{figure}
  \centerline{
    \includegraphics{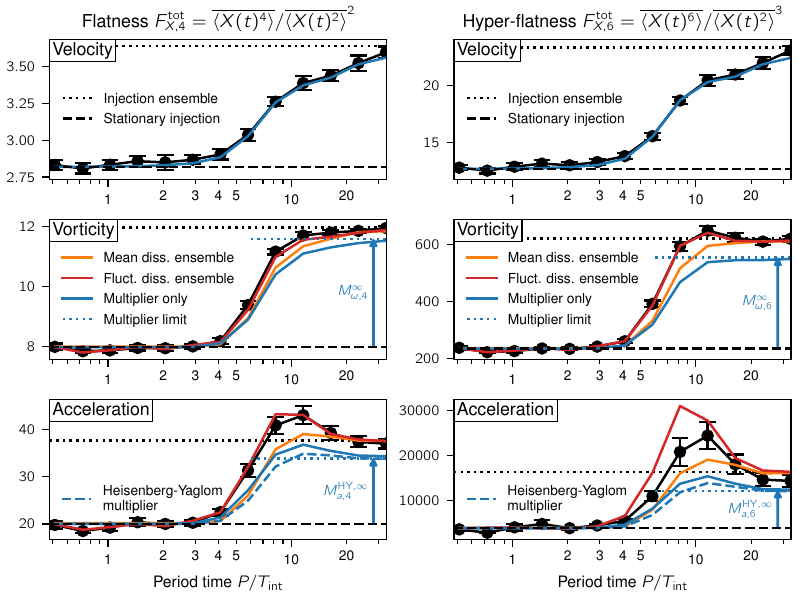}
  }
  \caption[width=\linewidth]{Flatness and hyper-flatness of velocity, vorticity, and acceleration at different period times of the oscillations (black dots and solid line). For small period times, the flatness converges to the one of the stationary reference simulation (black dashed line). For large period times, it converges to the injection ensemble (black dotted line), which superposes stationary turbulence statistics weighted according to the distribution of injection rates. The colored lines show different approximations of the flatness increase (based on $12\leq n \leq 256$ temporal bins per period). The error bars are standard errors obtained by assuming that moments computed from large temporal subintervals are i.i.d.\ Gaussian. These subintervals were chosen to be multiples of the period time of length at least $8\, T_\mathrm{int}$. The code and post-processed data used to generate this figure can be explored at \url{https://www.cambridge.org/S0022112024007006/JFM-Notebooks/files/Figure-4.ipynb}.
  \label{fig:flatness_prediction}}
\end{figure}

The flatness and hyper-flatness are shown in figure~\ref{fig:flatness_prediction} for varying period times (solid black line with dots). When changing the period time, the system transitions between two asymptotic behaviors (black dotted and dashed lines). At short period times, the oscillations are too fast for the flow to respond. Only the average energy input is relevant, which is the same as in the statistically stationary reference flow. Flatness values are the same as in this reference flow (black dashed line). For long period times, the system approaches the quasi-stationary behavior. Since the injection rate varies over time scales much longer than the dynamics of the flow, the flow has time to equilibrate to the instantaneous injection rate. Therefore, the instantaneous statistics are the same as given by a statistically stationary flow at the same injection rate. Time-aggregated statistics are then given by a superposition of the instantaneous flow statistics, weighted by the distribution of injection rates. This construction, which we call the `injection ensemble' (black dotted line), is detailed in section~\ref{sec:injensemble}. Finally, the transient regime between the two limits depends more crucially on the dynamical response of the flow to the changes in the forcing.

In the next section, we show that temporal mixing always increases flatness, with the main effect coming from mixing distributions with varying variance. In order to fully capture the transition of flatness, we then employ an ensemble approach. In a first step, we include the low-pass filter effect (see discussion of figure~\ref{fig:cascade_time} above). This leads to the ``mean dissipation ensemble'' (discussed in section~\ref{sec:injensemble}), which correctly captures both limiting behaviors. As we detail in sections~\ref{sec:dynamical_effects} and \ref{sec:fluct_diss_ens}, by additionally modeling the instantaneous flow states more precisely (``fluctuating dissipation ensemble''), we achieve good quantitative agreement in capturing the transition.
  
\section{Analysis and modeling}
\subsection{How temporal mixing increases flatness}
In order to understand the impact of temporal mixing, consider the instantaneous ($k$th-order hyper-)flatness
\begin{align} \label{eq:instant_flatness}
  F^\mathrm{inst}_{X,k}(t) &= \frac{\langle X(t)^k \rangle}{\langle X(t) ^2 \rangle^{k/2}}.
\end{align}
We will compare this instantaneous flatness to the total flatness $F_{X,k}^\text{tot}$. The difference is that the instantaneous flatness is computed for each phase of the oscillation, while for the total flatness, the averages are computed over all samples, aggregated over the phase of the oscillations. It is common practice to consider such time-averaged statistics in order to obtain large sample sizes.

\begin{figure}
  \centerline{
    \includegraphics{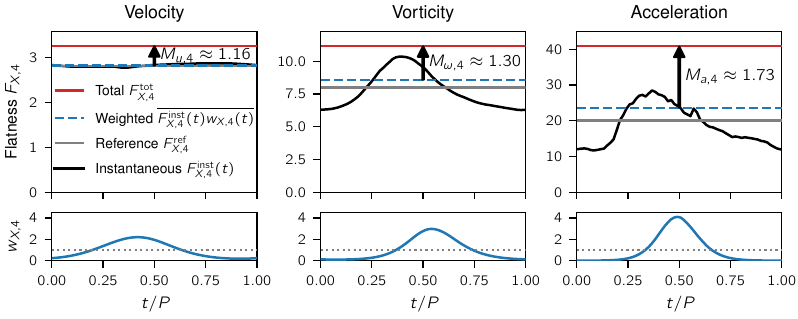}
  }
  \caption[width=\linewidth]{Instantaneous flatness with $k=4$ (upper panels, black solid) of velocity, vorticity, and acceleration for the simulation at $P \approx 8.2\,T_\mathrm{int}$. A grey solid line indicates the flatness of the reference simulation. The temporal weight (lower panels, blue solid) is used to compute the weighted average (upper panels, blue dashed), already resulting in a flatness value larger than the reference value of the stationary flow (grey solid). Additionally, the multiplier $M_{X,4}$ raises the flatness significantly (black arrows and red solid line).
  Note that the values depend slightly on the number of temporal bins, which here is $n=64$. The code and post-processed data used to generate this figure can be explored at \url{https://www.cambridge.org/S0022112024007006/JFM-Notebooks/files/Figure-5.ipynb}.\label{fig:flatness}}
\end{figure}

Figure~\ref{fig:flatness} compares the instantaneous flatness (black solid line) with the total flatness (red solid line) for the simulation at $P \approx 8.2\,T_\mathrm{int}$.
For the velocity, the instantaneous flatness does not vary notably within a period of the oscillations, staying slightly below the Gaussian value~3 (figure~\ref{fig:flatness}, upper left panel). Such sub-Gaussian velocity statistics are in line with more recent observations~\citep{jimenez1998,gotoh2002,mouri2002,WilczekJFM2011}.
The total flatness, however, takes a value above 3, purely as a result of the temporal mixing. Our results suggest that the difference between slightly sub-Gaussian and super-Gaussian velocity statistics can be attributed to large-scale flow variations. \citet{mouri2002} observed a transition from sub-Gaussian to super-Gaussian velocity statistics along decaying grid turbulence (accompanied by an increase of velocity gradient flatness), while the Reynolds number decreased. Our present observations corroborate their hypothesis that such behavior may be explained by an increasing degree of large-scale fluctuations further away from the grid.

For our vorticity and acceleration data, we observe notable variations of flatness in the course of a period (figure~\ref{fig:flatness}). As in the case of velocity, the total flatness surpasses any of the instantaneous values. This shows that the presence of large-scale fluctuations can significantly impact the flatness of these quantities, too. In the literature, acceleration flatness values scatter considerably, even for the same Reynolds numbers~\citep{VedulaPF1999,laporta2001,voth2002,mordant2004,BecJFM2006,YeungPF2006,ishihara2007,bentkamp2019}~(\citet{VedulaPF1999} report pressure gradient flatness). A possible explanation for this may be the differing degree of large-scale variations across simulations and experiments. In our simulations, we see that large-scale variations can indeed raise the acceleration flatness by roughly a factor of 2.

How does the additional averaging operation lead to this increase in flatness? We can rewrite the total flatness~\eqref{eq:total_flatness} as a function of the instantaneous flatness~\eqref{eq:instant_flatness} as
\begin{align}
  F_{X,k}^\text{tot} &= \frac{\overline{F^\mathrm{inst}_{X,k}(t) \langle X(t)^2 \rangle^{k/2}}}{\overline{\langle X(t)^2 \rangle}^{k/2}} \label{eq:raw_weight_multiplier}\\
  &= M_{X,k} \overline{w_{X,k}(t) F^\mathrm{inst}_{X,k}(t)}, \label{eq:weight_multiplier}
\end{align}
with the temporal weight
\begin{align}
  w_{X,k}(t) &= \frac{\langle X(t)^2 \rangle^{k/2}}{\overline{\langle X(t)^2 \rangle^{k/2}}} \ge 0, \quad \overline{w_{X,k}(t)} = 1,
\end{align}
and the multiplier
\begin{align} \label{eq:def_multiplier}
  M_{X,k} &= \frac{\overline{\langle X(t)^2 \rangle^{k/2}}}{\overline{\langle X(t)^2 \rangle}^{k/2}} \geq 1.
\end{align}
This means that the total flatness is a weighted time average of the instantaneous flatness, multiplied by a number $M_{X,k} \geq 1$. The inequality is a result of Jensen's inequality for powers, $\overline{f(t)^p} \geq \overline{f(t)}^p$, being valid for $p\geq 1$ and any non-negative time-dependent variable $f(t)$. Note that instead of time-dependent quantities, the statements about the flatness can be phrased analogously for any kind of conditional statistics such as considered in \citet{YeungPF2006}. Furthermore, similar expressions appear in superstatistical modeling~\citep[e.g.,][]{BeckPRE2005}.

Remarkably, this shows that temporal mixing always increases flatness. The only way to arrive at low flatness would be to have $w_{X,k}(t)$ emphasize periods of time where the instantaneous flatness is low, or to have no variations at all. In any case, an absolute lower bound is given by 
$F_{X,k}^\text{tot} \geq \min_t F^\mathrm{inst}_{X,k}(t)$. 
In our examples, however, the flatness lies significantly above this lower bound, even far above the value for the stationary flow. Based on \eqref{eq:weight_multiplier}, the increase can be decomposed into contributions from the multiplier and from the weighted average, as illustrated in figure~\ref{fig:flatness}.

In the remainder of this study, we investigate in more detail the mechanisms that lead to the total change in flatness as seen in figures~\ref{fig:flatness_prediction} and \ref{fig:flatness}. We discern three different effects, each leading to an amplification of flatness, which we  capture using increasingly complex models:
\begin{enumerate}
  \item \textit{Variations of variance.} Over time, the instantaneous variance of all considered quantities fluctuates. This automatically leads to an increase in flatness measured by the multiplier $M_{X,k}$. In the case of self-similar distributions, the instantaneous flatness becomes independent of time. As a result, \eqref{eq:weight_multiplier} reduces to $F_{X,k}^\text{tot} = M_{X,k}F^\mathrm{inst}_{X,k}$. So if the instantaneous distributions are self-similar (which is approximately the case for the velocity), then the temporal variations of the variance are the only effect leading to changes of flatness.
  Similarly, for vorticity and acceleration, variations of variance are a large contribution to flatness increase (see black arrows in figure~\ref{fig:flatness}).
  \item \textit{Mixing of non-self-similar statistics.} If the instantaneous flatness $F^\mathrm{inst}_{X,k}(t)$ changes over time, then the weighted average in \eqref{eq:weight_multiplier} will impact the flatness, too. To first approximation for slow variations, we can understand this as mixing of flow states that essentially look like stationary turbulence at different injection rates. So in order to understand the impact of the weighted average, we have to measure flatness and variance of stationary flows and superpose them in appropriate ways (we will call these `injection ensemble' and `mean dissipation ensemble').
  \item \textit{Dynamical effects.} Finally, for injection rate variations on time scales comparable to flow scales, there may be dynamical effects altering the instantaneous flatness values. For period times of approximately $6$ to $12$ integral times, we indeed observe such deviations from simple mixing, leading to even stronger instantaneous fluctuations and thus higher flatness. This will be modeled using the `fluctuating dissipation ensemble'.
\end{enumerate}
In the following, we will go through these effects in more detail.

\subsection{Variations of variance}
\label{sec:variationsofvariance}

The conceptually simplest contribution to the flatness increase is given by the multiplier $M_{X,k}$, generated by variations of variance. It is a dominant contribution for all quantities that we consider. 
Since the velocity PDF is in good approximation self-similar over time, its total flatness can be computed as
\begin{align}
  F_{u,k}^\mathrm{tot} \approx M_{u,k} F_{u,k}^\mathrm{ref} = \frac{\overline{\langle u^2(t) \rangle^{k/2}}}{\overline{\langle u^2(t) \rangle}^{k/2}} F_{u,k}^\mathrm{ref}= \frac{\overline{E(t)^{k/2}}}{\overline{E(t)}^{k/2}} F_{u,k}^\mathrm{ref}
\end{align}
from the periodically averaged time series of mean energy $E(t) = 3\langle u^2(t) \rangle/2$ and the flatness $F_{u,k}^\text{ref}$ of the stationary reference simulation. Here, we have assumed that the total velocity flatness of the stationary reference flow is the same as the (almost constant) instantaneous flatness in the oscillating flows. Figure~\ref{fig:flatness_prediction} shows that this multiplier indeed fully explains the increase of velocity flatness and hyper-flatness (blue line).

For vorticity and acceleration, the multiplier can serve as a first estimate of the flatness increase. Given the time series of the mean dissipation rate $\langle \varepsilon(t) \rangle$, the instantaneous vorticity variance is given exactly by $\langle \omega(t)^2 \rangle = \langle \varepsilon(t) \rangle / (3\nu)$ and the acceleration variance can be estimated by the Heisenberg-Yaglom prediction, $\langle a(t)^2 \rangle \approx a_0 \langle \varepsilon(t) \rangle^{3/2}\nu^{-1/2}$~\citep{Monin2013}. Approximating the weighted average in \eqref{eq:weight_multiplier} with the value from the stationary flow, we then have 
\begin{align}
  F_{\omega,k}^{\text{tot}} \approx M_{\omega,k} F_{\omega,k}^\text{ref} = \frac{\overline{\langle \varepsilon(t) \rangle^{k/2}}}{\overline{\langle \varepsilon(t) \rangle}^{k/2}} F_{\omega,k}^\text{ref} 
\end{align}
and
\begin{align} \label{eq:hy_estimate}
  F_{a,k}^{\text{tot}} \approx M_{a,k} F_{a,k}^\text{ref} \approx M_{a,k}^\text{HY} F_{a,k}^\text{ref} = \frac{\overline{\langle \varepsilon(t) \rangle^{3k/4}}}{\overline{\langle \varepsilon(t) \rangle^{3/2}}^{k/2}} F_{a,k}^\text{ref}.
\end{align}
Here, $F_{\omega,k}^\text{ref}$ and $F_{a,k}^\text{ref}$ denote the flatness values of the stationary reference simulation. In the case of the acceleration, the multiplier is further approximated as $M_{a,k}^\text{HY}$ using the Heisenberg-Yaglom prediction.
These estimates of the flatness capture a large part of the amplification but underestimate it (figure~\ref{fig:flatness_prediction}, blue lines). 
Nevertheless, they are a simple means to evaluate whether differences in flatness between two flows of comparable Reynolds number could be explained by large-scale intermittency.

For the sinusoidal modulations used in this study, the long-period limit of the vorticity and acceleration multipliers can even be computed analytically. In this limit, the periodically averaged dissipation rate $\langle \varepsilon(t) \rangle$ converges to the sinusoidal form of the injection rate $\xi(t)$. Therefore, we can write
\begin{align}
  M_{\omega, k} \xrightarrow{P\to\infty} M_{\omega, k}^\infty
  &= \frac{\overline{\xi(t)^{k/2}}}{\overline{\xi(t)}^{k/2}} 
  = \frac{1}{2\pi}\int_0^{2\pi}\dif t\,(1 + \alpha \sin t)^{k/2}\\
  &= 1 + 2\sum_{m=1}^{k/4} \binom{k/2}{2m} \binom{2m-1}{m} \left(\frac{\alpha}{2}\right)^{2m}\,,
\end{align}
where $\alpha=0.95$ is the relative amplitude of the modulations of the injection rate and $\binom{\cdot}{\cdot}$ denotes binomial coefficients. The sum is taken over all integers $1\leq m \leq k/4$. For the flatness, this amounts to $M_{\omega, 4}^\infty = 1+\frac{\alpha^2}{2} \approx 1.45$, and for the 6th-order hyper-flatness, we get $M_{\omega, 6}^\infty =  1 + \frac{3\alpha^2}{2} \approx 2.35$. Analogously, for the acceleration, evaluation of the integrals for the Heisenberg-Yaglom estimate, eq.~\eqref{eq:hy_estimate}, yields $M_{a,4}^\text{HY} \xrightarrow{} M_{a,4}^{\text{HY},\infty} \approx 1.69$ and $M_{a,6}^\text{HY} \xrightarrow{} M_{a,6}^{\text{HY},\infty} \approx 3.23$. These factors, multiplied with the value for the reference flow, are shown as blue dotted lines in figure~\ref{fig:flatness_prediction} and agree well with the data.

\subsection{Mixing of non-self-similar statistics}
\label{sec:injensemble}
In the previous subsection, we found that the main contribution to the flatness amplification comes from the multiplier. In order to get a more accurate estimate, however, we have to include effects of the weighted average in \eqref{eq:weight_multiplier}. In order to understand how this average leads to an increase in flatness, we here want to approximate the temporal mixing by the mixing of an ensemble of stationary turbulent flows. This is motivated by the limit of infinitely slow variations: if the variations of the injection rate are slower than any dynamics in the flow, then the flow will be statistically quasi-stationary, continually relaxing to a state of stationary turbulence at the current injection rate.

\begin{figure}
  \centerline{
    \includegraphics{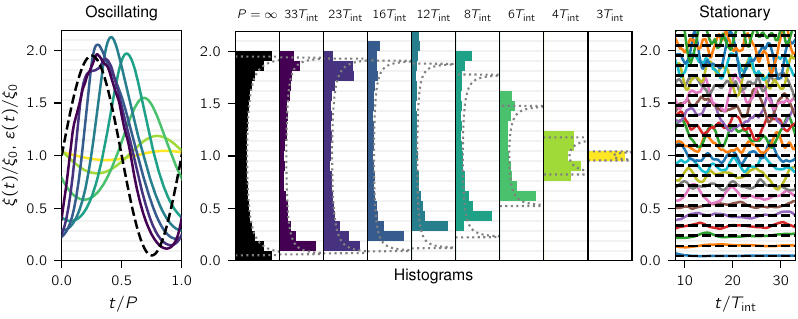}
  }
  \caption[width=\linewidth]{Visualization of how to construct the injection and mean dissipation ensembles. Left panel: injection rate time series (black dashed line) and periodically averaged dissipation rate time series (colored, solid lines) for the oscillating flows from $P\approx 33\,T_\mathrm{int}$ (violet) to $P\approx 3\,T_\mathrm{int}$ (yellow), using $69 \leq n \leq 256$ temporal bins. Due to the low-pass filtering of the energy cascade (see discussion of figure~\ref{fig:cascade_time}), the amplitude of the dissipation rate variations decreases with decreasing period times.
 Center panels: histograms of those periodically averaged curves. These histograms are used to construct the injection ensemble (leftmost histogram) and the mean dissipation ensembles (other histograms). Dotted lines indicate the distribution~\eqref{eq:injection_pdf} as a comparison, with rescaled $A_\xi$ to match the variance of the dissipation rate. Right panel: injection rates (black dashed lines) and dissipation rates (colored solid lines) of the statistically stationary ensemble members as functions of time. The code and post-processed data used to generate this figure can be explored at \url{https://www.cambridge.org/S0022112024007006/JFM-Notebooks/files/Figure-6.ipynb}. \label{fig:dissipation_histograms}}
\end{figure}

To assess this idea, we compare the statistics to an ensemble of 23 simulations with stationary energy injection rates (described in section~\ref{sec:dns}). Instead of mixing statistics in the temporal domain, we will mix statistics of the various statistically stationary flows. Figure~\ref{fig:dissipation_histograms} shows how the ensemble can be weighted so that its statistics match those of the various oscillating simulations.

The simplest way to do this is to match the distribution of injection rates. The sinusoidal injection rate~\eqref{eq:injection_rate} takes values distributed according to the PDF
\begin{align} \label{eq:injection_pdf}
  f(\xi) = \begin{cases}
    \frac{1}{\pi} \left[A_\xi^2 - \left(\xi - \xi_0\right)^2\right]^{-1/2}&\text{if } |\xi - \xi_0| < A_\xi, \\
    0 &\text{else.}
  \end{cases}
\end{align}
Weighting the ensemble members by this distribution gives rise to what we call the `injection ensemble'. As can be seen in figure~\ref{fig:flatness_prediction} (dotted black line), it accurately predicts the flatness in the limit of long period times. Since the distribution of injection rates~\eqref{eq:injection_pdf} does not depend on the period time, this ensemble cannot be used to capture the transition.

In order to capture the transition of vorticity and acceleration flatness, we introduce the `mean dissipation ensemble'. Instead of matching the distribution of injection rates, we here match the distribution of dissipation rates. To this end, we compute the mean dissipation rate for each temporal bin, then the distribution of these dissipation rates over the oscillation period. As shown in figure~\ref{fig:dissipation_histograms}, this leads to narrower ensemble weightings at short period times and to wider ensemble weightings at long period times due to the low-pass filter effect of the energy cascade. In the limit $P\to\infty$, the periodically averaged dissipation rate equals the injection rate, and the mean dissipation ensemble becomes identical to the injection ensemble described above (leftmost histogram in figure~\ref{fig:dissipation_histograms}). Accordingly, the flatness of the mean dissipation ensemble transitions between the value of stationary turbulence at short period times and the value of the injection ensemble at long period times (see figure~\ref{fig:flatness_prediction}, orange line). However, while giving the correct limits and better results than the pure multipliers, the mean dissipation ensemble still misses some of the flatness increase in the transition regime.

The mean dissipation ensemble is very much in the spirit of the K41 theory. It relies on the assumption that the instantaneous small-scale statistics in the form of vorticity and acceleration flatness are tied to the instantaneous value of the mean dissipation rate. The fact that it fails to predict flatness values correctly indicates that the small scales depend on more than just the average dissipation rate. This will be addressed in the next subsection, where we show that the combined values of dissipation mean and variance are needed to match statistics of the oscillating flows with ensembles of stationary turbulence.

\subsection{Dynamical effects} \label{sec:dynamical_effects}
\begin{figure}
  \centerline{
    \includegraphics{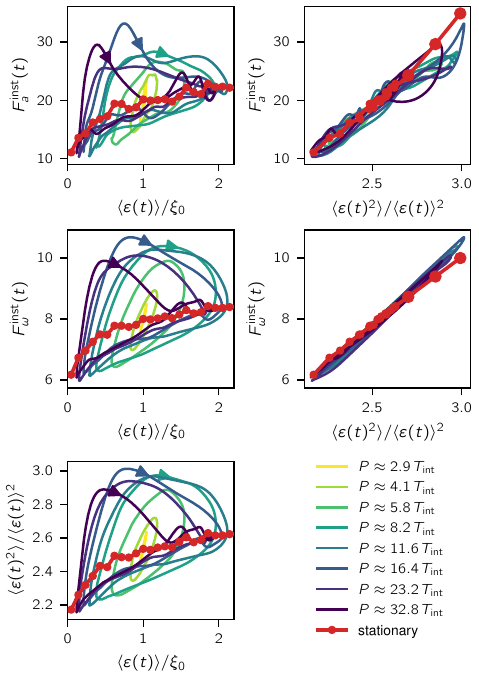}
  }
  \caption{Parametric plots of instantaneous statistics of the oscillating flows (moments periodically filtered with Gaussian filters of width $P/64$), compared with statistics of the statistically stationary flows (red line with dots). Left panels: For the oscillating simulations, flatness values of vorticity and acceleration as well as normalized second moment of dissipation behave similarly as a function of the mean dissipation. Right panels: All flatness curves approximately collapse when plotting them as a function of the normalized second moment of dissipation. Note that we have added simulations of stationary turbulence at $\xi(t) = \{4, 9, 18\}\xi_0$ to extend the red lines (larger dots). The code and post-processed data used to generate this figure can be explored at \url{https://www.cambridge.org/S0022112024007006/JFM-Notebooks/files/Figure-7.ipynb}. \label{fig:loop_plots}}
\end{figure}
For period times of the order of 6 to 12 integral times, figure~\ref{fig:flatness_prediction} still shows notable deviations between vorticity and acceleration flatness of the mean dissipation ensemble compared to the oscillating flows. According to \eqref{eq:raw_weight_multiplier}, this must be due to a mismatch in instantaneous variance or flatness. Based on how we constructed the ensemble, it means that the instantaneous statistics of vorticity and acceleration are not uniquely related to the instantaneous mean dissipation rate when comparing between the oscillating flows and stationary turbulence.

This is investigated more closely in figure~\ref{fig:loop_plots}, showing how the various quantities are related by parametric plots, both in the oscillating case (yellow to violet) and in the stationary case (red with dots, each dot representing one simulation). In addition to the vorticity and acceleration flatness, we show the normalized second moment of dissipation, $\langle \varepsilon(t)^2 \rangle/\langle \varepsilon(t) \rangle^2$, as a measure of the fluctuations around the mean dissipation rate~\citep[see also][]{GaudingPotCI2021}. None of these three quantities appears to be determined uniquely by the instantaneous mean dissipation rate $\langle\varepsilon(t)\rangle$ (figure~\ref{fig:loop_plots}, left panels). For very slow, quasi-static changes of the injection rate (infinite period time), we would expect the oscillating curves (yellow to violet) to collapse onto the one for stationary turbulence (red with dots). However, indications of this become visible only at the longest period times. Instead, for all curves we observe a larger flatness on their intensifying branch (increasing dissipation rate) than on their decaying branch (decreasing dissipation rate), most pronouncedly at intermediate period times such as $P \approx 11.6\,T_\mathrm{int}$. This indicates that intensifying turbulence is characterized by heavier-tailed dissipation, vorticity, and acceleration statistics than its stationary equivalent.

However, plotting vorticity and acceleration flatness against $\langle \varepsilon(t)^2 \rangle/\langle \varepsilon(t) \rangle^2$ (right panels) shows that the excursions seen in the left panels share some similarities. Curves for different period times approximately collapse, which means that vorticity and acceleration flatness and normalized second moment of dissipation scale in the same way, independent of the oscillation frequency. In fact, they even behave similarly to the scaling of stationary turbulence (red line with dots). Note that in order for the red lines to reach this far, the base ensemble was extended by three stationary turbulence simulations on a $1024^3$ grid with significantly higher injection rates (larger red dots; for more details see section~\ref{sec:dns}).

\subsection{Fluctuating dissipation ensemble} \label{sec:fluct_diss_ens}
\begin{figure}
  \centerline{
    \includegraphics{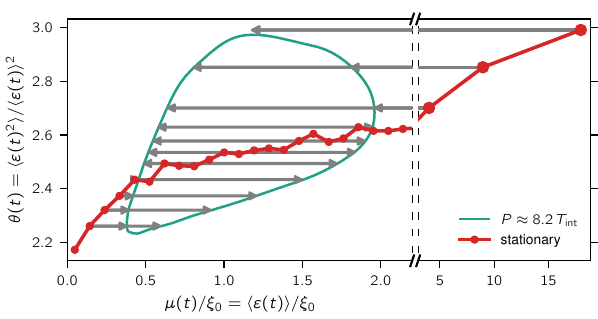}
  }
  \caption{Visualization of the procedure to construct the fluctuating dissipation ensemble for the oscillating flow at $P \approx 8.2\,T_\mathrm{int}$. The green and red curves are the same as in the lower left panel of figure~\ref{fig:loop_plots}. For each instantaneous value $\theta(t)$ of the oscillating flow, the stationary flow (red dots) with the closest value of $\theta$ is selected. Then it is rescaled at constant Reynolds number and viscosity, such that also the mean dissipation rate $\mu$ is matched (grey arrows). The superposition of the statistics of all the rescaled stationary flows selected in this way then constitutes the fluctuating dissipation ensemble. The code and post-processed data used to generate this figure can be explored at \url{https://www.cambridge.org/S0022112024007006/JFM-Notebooks/files/Figure-8.ipynb}.
   \label{fig:loop_schematic}}
\end{figure}
We observed that the instantaneous flows display features of higher-Reynolds-number turbulence, manifesting in large values of the instantaneous flatness and the normalized second moment of dissipation. Based on this, we can construct a refined ensemble model, which we call `fluctuating dissipation ensemble'. For each instantaneous value of $\theta(t) = \langle \varepsilon(t)^2 \rangle/\langle \varepsilon(t) \rangle^2$, we select a simulation of stationary turbulence that matches this value most closely (see figure~\ref{fig:loop_schematic}). Then we know, based on the approximate collapse in figure~\ref{fig:loop_plots} (right panels), that the selected stationary flow will approximate the instantaneous values of vorticity and acceleration flatness, too.
However, we know that total flatness is determined not only by instantaneous flatness but also by instantaneous variance (see \eqref{eq:weight_multiplier}). Hence we do not expect this ensemble model to predict the total flatness well, yet.

In order to extend the set of simulations of stationary turbulence that we have, we perform Reynolds-similar rescalings of the data. This is justified by the fact that every solution of the Navier-Stokes equation such as our simulation data can be rescaled in space and time and will still be a solution of the Navier-Stokes equation~\citep{Frisch1995}. The rescaling factors in space, $\alpha_S$, and in time, $\alpha_T$, can be chosen freely. In the rescaled system, each quantity is given by its original value multiplied with the rescaling factors according to its units, e.g., the rescaled dissipation rate is given by
\begin{align}
  \varepsilon' = \frac{\alpha_S^2}{\alpha_T^3} \varepsilon.
\end{align}
Dimensionless quantities such as the flatness and the Reynolds number remain constant. This method allows us to generate a family of different flows from a single simulation run, connected by Reynolds similarity. Since we have two degrees of freedom for the rescaling, we can decide to keep the viscosity (units $\mathrm{Length}^2/\mathrm{Time}$) constant by choosing $\alpha_S^2 = \alpha_T \equiv \alpha^2$. The data that we generate by this method are the same as what would be obtained by running a simulation at the same Reynolds number and viscosity but with different forcing scale and injection rate.

Let us now use this method to construct our advanced ensemble model. For the mean dissipation ensemble, we matched the distribution of mean dissipation rate $\mu(t)=\langle \varepsilon(t)\rangle$ between the oscillating flow and the ensemble. Now we are matching the distribution of $\theta(t) = \langle \varepsilon(t)^2 \rangle/\langle \varepsilon(t) \rangle^2$. However, in order to get a good quantitative match between ensemble model and oscillating flow, it would be better to match both quantities at the same time, i.e. the full PDF of $\mu$ and $\theta$,
\begin{align}
  f(\mu, \theta) = \overline{\delta\left(\mu - \langle \varepsilon(t) \rangle\right) \delta\left(\theta - \tfrac{\langle \varepsilon(t)^2 \rangle}{\langle \varepsilon(t) \rangle^2} \right)}.
\end{align}
Using the rescaling approach, this is done in the following way: For a given combination of $\mu$ and $\theta$, we select the stationary flow that matches the value of $\theta$ most closely (see figure~\ref{fig:loop_schematic}). In general, the mean dissipation rate of the $\theta$-matched simulation $\langle \varepsilon \rangle_\theta$ will not be equal to $\mu$. We can, however, choose a rescaling factor $\alpha$ such that 
\begin{align}
  \mu = \frac{\alpha_S^2}{\alpha_T^3} \langle \varepsilon \rangle_\theta
  = \alpha^{-4} \langle \varepsilon \rangle_\theta. \label{eq:rescaling}
\end{align}
After the rescaling, both $\mu$ and $\theta$ and thus both mean and variance of the dissipation rate are matched (figure~\ref{fig:loop_schematic}, grey arrows). Based on the observations in this study, we can hope that this will allow us to approximate both lower- and higher-order instantaneous statistics of vorticity and acceleration, too.

Finally, we superpose the statistics of the rescaled statistically stationary flows to imitate the time-averaged data of the oscillating flow. For example, for the acceleration PDF, which rescales with $\alpha_S/\alpha_T^2= \alpha^{-3}$, we have
\begin{align}
  f^\mathrm{ens}(a) = \int \dif \mu \int \dif \theta f(\mu, \theta) \alpha(\mu, \theta)^{-3}f^\mathrm{stat}_\theta(a \alpha(\mu, \theta)^3).
\end{align}
Here, $f^\mathrm{ens}$ denotes the acceleration PDF of the ensemble, and $f^\mathrm{stat}_\theta$ denotes the acceleration PDF (before rescaling) of the stationary flow with the given value of $\theta$. The rescaling factor $\alpha(\mu, \theta)$ can be computed from $\mu$ and $\theta$ by \eqref{eq:rescaling}. Any other flow statistics can be computed likewise. In particular, figure~\ref{fig:flatness_prediction} shows vorticity and acceleration flatness and hyper-flatness values of this ensemble (red lines).

The fluctuating dissipation ensemble excellently captures vorticity flatness and 6th-order hyper-flatness (see figure~\ref{fig:flatness_prediction}). In particular, it successfully reproduces the increase in flatness due to dynamical effects that was missing from the mean dissipation ensemble. Although we incorporated only the first two moments of dissipation in the formulation of the model (which are related to second and fourth moments of vorticity), the 6th-order hyper-flatness of vorticity is matched perfectly, too. This indicates that a simple measure of dissipation fluctuations such as $\langle \varepsilon(t)^2 \rangle/\langle \varepsilon(t) \rangle^2$ may be sufficient for accurate modeling of even higher-order small-scale statistics. For acceleration statistics, flatness is captured quite well, while there are deviations for the hyper-flatness. Some of the inaccuracies of the model can already be surmised from the slight mismatch of flatness between oscillating and stationary flows in the right panels of figure~\ref{fig:loop_plots}. Overall, the ensemble models are well suited to capture amplification of higher-order moments due to large-scale intermittency. When the large-scale fluctuations are happening on time scales comparable to a few integral scales of the flow, it becomes crucial to include also the second moment of the dissipation rate into the modeling.

\section{Comparison to higher Reynolds number} \label{sec:higher_Re}
\begin{figure}
  \centerline{
    \includegraphics{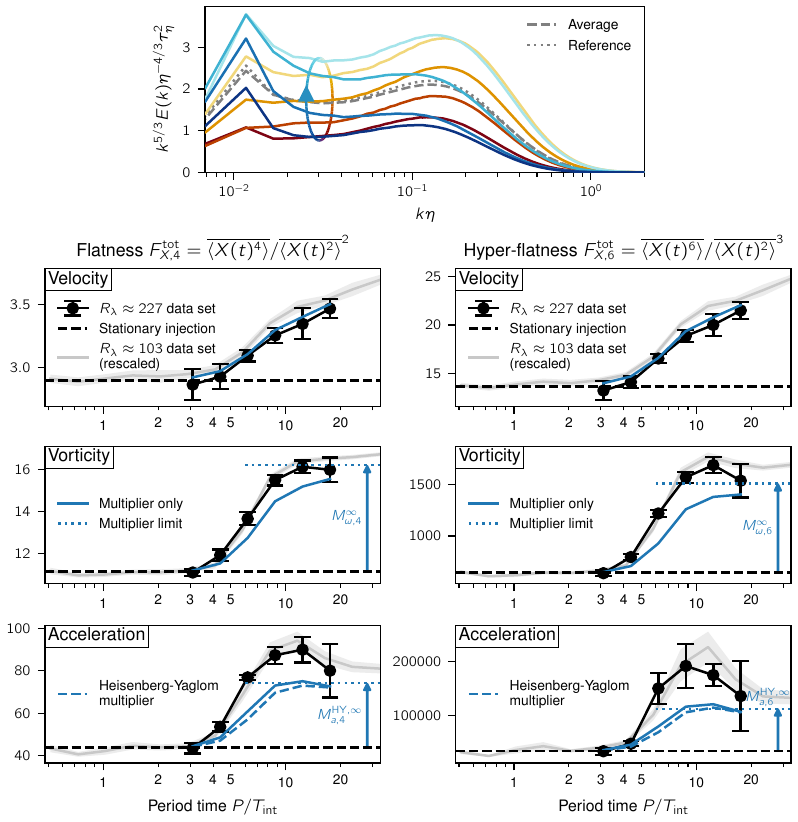}
  }
  \caption{Top: Periodically averaged compensated energy spectra for the oscillating flow with period time $P\approx 8.8\,T_\mathrm{int}$ in the higher Reynolds number data set. The bins are the same as used in figure~\ref{fig:breathing_pdfs}. Bottom: Flatness and hyper-flatness of velocity, vorticity, and acceleration as functions of period time for the oscillating flows in the higher Reynolds number data set (black dots and solid line). Error bars are computed as in figure~\ref{fig:flatness_prediction}. The reference flow with stationary injection rate at $R_\lambda^\mathrm{ref} \approx 227$ (black dashed line) corresponds to the limit of short period times. Multipliers are shown as blue lines as in figure~\ref{fig:flatness_prediction}, computed using $147 \leq n \leq 256$ temporal bins per period. Ensemble models are not available for this data set. The flatness of the oscillating flows from the main data set (as in figure~\ref{fig:flatness_prediction}, black dots and solid line) is shown in grey in the background, multiplied by the ratio of the flatness values of the respective reference flows. The code and post-processed data used to generate this figure can be explored at \url{https://www.cambridge.org/S0022112024007006/JFM-Notebooks/files/Figure-9.ipynb}.
   \label{fig:higherRe}}
\end{figure}
Achieving good convergence for the high-order moments of the oscillating flows requires very long simulation times, making the full analysis including the ensemble modeling feasible only for simulations on $512^3$ grid points. In order to evaluate the Reynolds number dependence of our results, we additionally simulated six oscillating flows on a $1024^3$ grid, accompanied by a reference flow with constant injection rate at $R^\mathrm{ref}_\lambda \approx 227$ and spatial resolution $k_M \eta \approx 2.0$. The duration of the simulations is between $44\,T_\mathrm{int}$ and $88\,T_\mathrm{int}$ (with $T_\mathrm{int}$ measured in the reference flow). The simulation parameters are the same as described for the main data set in section~\ref{sec:dns}, only at smaller viscosity. The period times $P$ are varied between $3.1\,T_\mathrm{int}$ and $17.5\,T_\mathrm{int}$. The spatial resolution of the oscillating flows varies between $1.7 \leq k_M\eta \leq 3.1$. Due to the computational constraints, we did not consider any ensemble of stationary flows in this case.

The top panel of figure~\ref{fig:higherRe} shows the periodically averaged energy spectrum of the flow at $P\approx 8.8\,T_\mathrm{int}$, which can be compared directly to that shown in figure~\ref{fig:breathing_pdfs}. While the overall dynamics looks very similar, at the higher Reynolds number we can observe a more pronounced inertial range. The bottom panels are analogous to figure~\ref{fig:flatness_prediction}. They show how the flatness and hyper-flatness increase as a function of the period time of the oscillations. As a direct comparison, we show the main data from figure~\ref{fig:flatness_prediction} as a grey line in the background. In order to make them comparable, we divided those data by the values from the reference flow of the main data set and multiplied them by the values of the reference flow from the higher-Reynolds-number data set.

Given that we compensate the data only by the respective integral time scales $T_\mathrm{int}$ and reference flatness values $F_{X,k}^\mathrm{ref}$, the increase in flatness is remarkably similar between the two Reynolds numbers, within the error bars of our data. In particular, the factor by which the flatness increases is very close. This can be explained by the multiplier mechanism detailed in section~\ref{sec:variationsofvariance}. There, we showed that we can compute exactly the infinite-period-time limit of the vorticity multiplier and of the Heisenberg-Yaglom estimate of the acceleration multiplier, and we found that they are independent of Reynolds number. As in figure~\ref{fig:flatness_prediction}, these are shown as blue dotted lines. Since the multipliers account for a large part of the (hyper-)flatness increase, it is expected that we observe a comparable increase at different Reynolds numbers. While the remaining effect of the weighted average in eq.~\eqref{eq:weight_multiplier} may have non-trivial Reynolds number dependence, these are good reasons to believe that the effect of large-scale intermittency will be of similar magnitude at even higher Reynolds numbers.

Furthermore, it is very interesting that the critical period time of the transition (in units of $T_\mathrm{int}$) is almost identical at this higher Reynolds number. As described in the discussion of figure~\ref{fig:cascade_time}, we found a low-pass filter effect of turbulence, which dampens the effect of large-scale intermittency at shorter period times. The fact that we observe a collapse at two different Reynolds numbers shows that it is mainly the integral flow scales determining the time scale of this filter. 

We conclude this section by discussing some practical implications. Overall, the preceding discussion indicates that both the time scales of the transition (in integral units) and the magnitudes of the multipliers will remain comparable even at higher Reynolds number. This means that it is possible to estimate the magnitude of the effect of large-scale intermittency on higher-order moments by measuring the amplitude and time scales of the large-scale variations in a turbulent flow. If the frequency of the variations is comparable to or lower than $\sim 1/(5T_\mathrm{int})$, then it is likely that the flatness measurements are affected by them. Moreover, the factor by which the flatness is likely increased due to large-scale intermittency can be estimated by a computation such as the one detailed in section~\ref{sec:variationsofvariance}.

\section{Conclusion}
The idea of universality in turbulence relies on the assertion that the smallest scales of the flow are the result of its internal nonlinear dynamics and therefore become independent of the external forcing mechanism that generates the turbulence. Next to the Reynolds number, it was put forward by Landau~\citep{Landau2013} that sufficiently slow, large-scale variations will constrain universality. In order to ensure comparability across different turbulent flows, it is therefore important to identify and characterize the properties of the flow that affect small-scale universality. 
Here, we evaluated the relevance of Landau's remark for the statistics of velocity, vorticity, and acceleration.

We computed the flatness and 6th-order hyper-flatness of these quantities in direct numerical simulations of unsteady, homogeneous, isotropic turbulence subject to sinusoidal variations of the energy injection rate with different period times. Analytically, we found that temporal mixing always increases flatness. For the velocity, large-scale fluctuations can make the difference between sub-Gaussian and super-Gaussian statistics. For the acceleration, we observed a flatness increase by approximately a factor of~2.

We identified three different mechanisms that each lead to an increase in flatness. First, any variations of variance over time lead directly to an increase in flatness that can be measured by the multiplier $M_{X,k}$. This fully explains the increase of velocity flatness. The multipliers can be either computed directly or approximated from the time series of mean energy and dissipation rate. Second, through weighted averaging, flatness increases when large instantaneous flatness values coincide with large variance. Together, these two effects can be modeled by an appropriate superposition of statistics of stationary turbulence, identifying each instantaneous state of the oscillating flows with a flow of stationary turbulence based on the mean dissipation rate. The third effect occurs only for period times of the order of 6 to 12 integral times. In that regime, we observed that the fluctuating input generates stronger excursions of the instantaneous flatness than expected from the identification with stationary turbulence. We found that this is also reflected in the normalized second moment of dissipation, which can act as a good predictor for flatness of both vorticity and acceleration. Within the scope of the quantities considered here, we found that dissipation mean and variance together are a good representation of the instantaneous statistical state of the small scales, and that they can be used to construct an ensemble model that accurately captures both regular flatness and 6th-order hyper-flatness.

While we focused here on the flatness of velocity, vorticity, and acceleration, we put forward ensemble descriptions that capture the full flow statistics at once. Since these models combine the data from simpler turbulent flows to model more complex turbulent flows, they require only assumptions about how to superpose these statistics. Meanwhile, they can make predictions about many different flow statistics. An interesting direction of future work will be to investigate whether similar ensembles of lower-Reynolds-number flows can be used to model higher-Reynolds-number turbulence. 

We introduced here artificial variations of the energy injection rate into statistically homogeneous and isotropic flows and found that these variations had a strong effect on higher-order statistics. With this work, we hope to spark interest in further studies on large-scale intermittency, including investigations of the strength and impact of `naturally occurring' large-scale fluctuations in experimental and in other simulated flows. Quantifying their effect as a correction to the Reynolds number effects may be crucial to explain the observed scatter of higher-order statistics in the literature. This may be achieved by accompanying reports of higher-order statistics by quantities such as the multipliers used in our study or coefficients of inertial-range scaling laws as proposed by \citet{chien2013}. If such an analysis allows us to match data across different flow types more precisely, then this would make a strong point for a formulation of universality explicitly conditional on large-scale variations.
Overall, we hope that this work can contribute to improve comparability in turbulence by better characterizing the effect of large-scale intermittency.

\backsection[Supplementary material]{
Supplementary movies and computational notebook files are available at \url{https://doi.org/10.1017/jfm.2024.700}. The simulations were conducted using our code TurTLE~\citep{LalescuCPC2022}, which is available at \url{https://gitlab.mpcdf.mpg.de/TurTLE/turtle}. The post-processing code and the post-processed data can be accessed at \url{https://www.cambridge.org/S0022112024007006/JFM-Notebooks} along with interactive notebooks generating each of the figures in this study.}

\backsection[Acknowledgements]{
  We would like to acknowledge interesting and useful discussions with Maurizio Carbone and Greg A.\ Voth as well as technical support from Cristian C.\ Lalescu. We thank the anonymous reviewers for their constructive comments, which helped to improve the manuscript. GitHub Copilot was used during code development.
}

\backsection[Funding]{
  This project has received funding from the European Research Council (ERC) under the
  European Union's Horizon 2020
  research and innovation programme (Grant agreement No.\ 101001081).
  
  The authors gratefully acknowledge the scientific support and HPC resources provided by the Erlangen National High Performance Computing Center (NHR@FAU) of the Friedrich-Alexander-Universit\"at Erlangen-N\"urnberg (FAU) under the NHR project \mbox{``EnSimTurb''}. NHR funding is provided by federal and Bavarian state authorities. NHR@FAU hardware is partially funded by the German Research Foundation (DFG) -- 440719683.}

\backsection[Declaration of interests]{The authors report no conflict of interest.}

\backsection[Author ORCIDs]{\newline
  L.\ Bentkamp, \url{https://orcid.org/0000-0001-6798-9229};\newline
  M.\ Wilczek, \url{https://orcid.org/0000-0002-1423-8285}}

\backsection[Author contributions]{M.W. designed the study. L.B. carried out the numerical simulations and analysis. Both authors analyzed the data and wrote the manuscript.}

\bibliographystyle{jfm}
\bibliography{refs}

\end{document}